\documentclass[runningheads]{llncs}

\usepackage{amsmath}
\usepackage{amssymb}
\usepackage{graphicx}
\usepackage{booktabs}
\usepackage[numbers,sort&compress]{natbib}
\usepackage[normalem]{ulem}
\usepackage{xcolor}
\usepackage{pdfpages}
\usepackage[switch]{lineno}
\usepackage[colorlinks]{hyperref}
\usepackage{xurl}  % allow URLs to break anywhere; clicked target remains the full URL

\newcommand{\appendixref}[1]{\hyperref[#1]{Appendix~\ref*{#1}}}

\newcommand{\bE}{{\mathbb{E}}}
\newcommand{\cF}{{\mathcal{F}}}
\newcommand{\cW}{{\mathcal{W}}}
\newcommand{\cS}{{\mathcal{S}}}
\newcommand{\cC}{{\mathcal{C}}}
\newcommand{\cR}{{\mathcal{R}}}

\title{Risk-Limiting Audits for Parliamentary Majorities\thanks{Authors listed
alphabetically. Accepted for \href{https://e-vote-id.org/}{E-Vote-ID 2026}.
This work was supported by the Australian Research Council
(OPTIMA ITTC IC200100009).}}

\author{
Jack Freestone   \inst{1} \orcidID{0009-0008-2983-6676} \and
Dennis Leung     \inst{2} \orcidID{0000-0002-5924-8161} \and
Damjan Vukcevic  \inst{3} \orcidID{0000-0001-7780-9586}
}
\authorrunning{Freestone J, Leung D, Vukcevic D}

\institute{
School of Mathematical and Physical Sciences, Macquarie University, %Sydney,
Australia
\and
School of Mathematics and Statistics, University of Melbourne, %Parkville,
Australia
\and
Department of Econometrics and Business Statistics, Monash University, %Clayton,
Australia
\\
\email{damjan.vukcevic@monash.edu}
}

\begin{document}

\maketitle

\begin{abstract}
Existing methods for risk-limiting audits typically focus on certifying
individual contests. In parliamentary elections, however, the politically
relevant outcome is often whether a party has won enough seats to form
government, not whether every reported seat outcome is correct.  Extending
on the work of \citet{mohanty2019auditing}, we formulate the certification
of a parliamentary majority as a partial conjunction testing problem: it is
enough to verify that the reported winning party truly won at least a
majority of its reported seats.  Building on the SHANGRLA auditing framework,
we construct a sequential audit statistic for the majority outcome by
combining seat-level statistics. We then propose adaptive sampling strategies
that allocate auditing effort across seats, including variants that learn to
avoid spending excessive effort on seats that appear unlikely to have been
truly won.  Using simulations based on synthetic and real data, from the
2014 Indian Lok Sabha election, we show that auditing the parliamentary
majority can substantially reduce the number of ballots inspected (by almost
a thousand-fold) compared to certifying every reported winning seat.
\end{abstract}
%\keywords{}

%% ================================================================
\section{Introduction}
\label{sec:intro}
%% ================================================================

A \emph{risk-limiting audit} (RLA) is a post-election procedure that
sequentially samples ballots, accumulating statistical evidence until either
the reported outcome can be certified\footnote{%
We use `certify' (and `certification') throughout in a purely
\emph{statistical} sense: to verify the result of the election with risk limit
$\alpha$.  This is distinct from \emph{certification} as a \emph{legal} act,
whose meaning vary across jurisdictions.%
} or the audit escalates to a full hand count \citep{lindeman2012gentle}. Its
guarantee is that if the reported outcome is wrong, the audit certifies it with
probability at most~$\alpha$ (the \emph{risk limit}), a pre-specified value. If
the reported outcome is correct, certification typically needs only a small
fraction of the ballots. The SHANGRLA framework of \citet{stark2020sets}
reduces the auditing of any electoral system (a \emph{social choice function})
to testing a finite collection of \emph{assertions} about population means; the
ALPHA \emph{test supermartingale} \citep{stark2023alpha} provides a flexible
sequential test for each. These tools, however, were developed primarily for
single-contest elections. In a parliamentary system, government is determined
not by any single race but by whether a party controls a majority of seats, a
joint outcome whose auditing has received comparatively little attention.

As first observed by \citet{mohanty2019auditing}, certifying a parliamentary
majority does not require certifying every individual seat.  We formalise this
as a \emph{partial conjunction} hypothesis test \citep{benjamini2008screening}:
a procedure that rejects at least some (but not necessarily all) null
hypotheses from a large collection. This pools evidence across seats,
certifying the majority even when some seats have not individually accumulated
enough evidence, and can substantially reduce the number of ballots that must
be hand-inspected.

We develop a suite of adaptive sampling schemes that dynamically allocate
auditing effort across seats. These include \emph{greedy} strategies that
concentrate sampling on weaker seats (where stronger evidence is required) and
\emph{filtering} strategies that redirect effort away from seats where the
reported outcome seems to be false.
We evaluate our schemes via simulation across a range of parliamentary
configurations, including scenarios based on the 2014 Indian Lok Sabha
election, and show they require observing orders of magnitude fewer ballots
than audits that aim to certify all reported winning seats.

%% ================================================================
\section{RLA for a parliamentary majority by sequentially testing a partial
conjunction hypothesis}
\label{sec:methods}
%% ================================================================

Let $\cS$ denote the set of all seats in a parliament, each corresponding to a
separate contest.  As a concrete example, the Lok Sabha (Lower House) of the
Indian Parliament has $|\cS| = 543$ seats corresponding to 543 constituencies
\citep{mohanty2019auditing}. Whichever party controls the majority, that is,
\[
r \equiv \left\lfloor |\cS|/2 \right\rfloor + 1
\]
of the $|\cS|$ seats, can form the government for the next term. Suppose $\cW
\subset \cS$ is the subset of seats reportedly won by the winning party
after a given election (necessarily $|\cW| \geqslant r$). To confirm that the
new government can be rightfully formed, we need to verify that at least $r$
of the seats in $\cW$ have been truly won by the winning party, which can be
formulated as testing the null hypothesis
\begin{equation*} \label{PC_null_for_majority}
H_{r/\cW}\colon
\text{fewer than $r$ seats in $\cW$ have been truly won by the winning party}.
\end{equation*}
This is the complementary null of the assertion that the winning party won at
least $r$ seats.
If $H_{r/\cW}$ is rejected, the majority outcome is considered certified.

To conduct an audit with risk limit $\alpha$, we can sample the ballots cast
for the seats in $\cW$ sequentially, until their revealed evidence suggests
that $H_{r/\cW}$ can be rejected at level $\alpha$.  We conceptualise such
sampling as happening iteratively in `rounds', indexed by a time variable $t$.
Before any sampling has taken place we have $t = 0$, and at each subsequent
time $t \in \{1, 2, \dots\}$, the auditor has the option to draw one cast
ballot for each seat $s \in \cW$ without replacement.\footnote{%
This choice is primarily for clear exposition.  In practice, a real audit would
aim to have very few rounds, and would draw a much larger sample of ballots in
each round.  The methods we develop here are applicable to any such sampling
scheme.}
Let
\[
D_{s, t + 1} \equiv \begin{cases}
1, & \text{if a ballot for seat $s$ is drawn at time $t + 1$},\\
0, & \text{otherwise},
\end{cases}
\]
be the auditor's sampling decision for time $t + 1$ for seat $s$.
The decision $D_{s, t + 1}$ can therefore only depend on the information from
the ballots sampled up to time $t$; we denote this information as
$\mathcal{F}_t$.

To evaluate the evidence against a given null hypothesis, $H$, we need to
define a suitable test statistic, $U_t$.  We will calculate this iteratively
based on the information revealed in $\cF_t$, giving rise to a sequence of
values, $U \equiv (U_t)_{t \geqslant 0}$, referred to as a \emph{(test)
process}.  We want processes that satisify the following:
\begin{enumerate}
\item
We say that $U$ is \emph{anytime-valid for testing $H$} if
\begin{equation} \label{anytime_validity_generic}
\Pr\left(\sup_{t \geqslant 0} U_t  \geqslant 1/\alpha\right) \leqslant \alpha
    \text{ for all } \alpha \in (0, 1) \text{ when $H$ is true}.
\end{equation}
This allows us to reject $H$, with risk limit $\alpha$, once $U$ exceeds
$1 / \alpha$.
\item
If $H$ is false, we want to reject it efficiently.
We seek a process $U$ that will grow rapidly in response to new observations,
so it can quickly exceed $1 / \alpha$.
\end{enumerate}
In the remainder of this section we construct a process ($E_{r/\cW} $ in
\autoref{thm:main}) that is anytime-valid for testing $H_{r/\cW}$. Then in
\autoref{sec:sampling_schemes} we explore various sampling schemes that guide
the drawing decisions $D_{s, t}$ across time, aiming to minimize the total
number of ballots sampled before $H_{r/\cW}$ can be rejected.

\subsection{A test process for certifying a single seat}
\label{sec:testing_one_seat}

As a starting point, we define for each seat $s \in \cW$ the null hypothesis
\begin{equation*} \label{seat_level_hypothesis}
H_s\colon \text{the winning party has \emph{not} truly won the seat $s$},
\end{equation*}
and let $N_s$ be the total number of ballots cast for seat~$s$, which we
label $b_{s, 1}, \dots, b_{s, N_s}$.
We assume the social choice function for this seat can be decomposed into a
finite set of $J_s$ \emph{assertions} using the SHANGRLA framework
\citep{stark2020sets}, i.e., there exist $J_s$ non-negative \emph{assorter}
functions $\{A_{s, j}\}_{j = 1}^{J_s}$ such that, by letting $x_{s, j, i}
\equiv A_{s, j}(b_{s, i})$, $H_s$ is false if and only if
\begin{equation} \label{eqn:assertions}
\bar{x}_{s,j} \equiv \frac{1}{N_s} \sum_{i=1}^{N_s} x_{s,j,i} > 1/2
    \quad \text{for all } j = 1, \dots, J_s.
\end{equation}
We can therefore write
\begin{equation} \label{union_null}
H_s = \bigcup_{j = 1}^{J_s} H_{s, j},
\end{equation}
where
\[
H_{s, j}\colon \bar{x}_{s,j} \leqslant 1/2
\]
is the null hypothesis that the $j$th assertion is \emph{not} true.  As an
example, consider a seat contested by just two candidates, Alice (the reported
winner) and Bob. Here there is a single assertion ($J_s = 1$), whose assorter
$A_{s,1}$ scores a ballot for Alice as~$1$, a ballot for Bob as~$0$, and any
other (e.g.\ invalid) ballot as~$1/2$. The assertion mean $\bar{x}_{s,1}$ is
then the average of these scores across all $N_s$ ballots, which exceeds~$1/2$
if and only if Alice received more votes than Bob.

Suppose $(B_{s, i})_{i = 1}^{N_s}$ represents the full sequence of ballots that
would be drawn randomly without replacement for seat $s$, which is necessarily
a random permutation of $\{b_{s, i}\}_{i = 1}^{N_s}$.
Let $(X_{s, j, i})_{i = 1}^{N_s} = (A_{s, j}(B_{s, i}))_{i = 1}^{N_s}$
represent the corresponding (random) sequence of values from
$\{x_{s, j, i}\}_{i=1}^{N_s}$. Since a ballot for seat $s$ isn't necessarily
drawn in every round, one should keep in mind that $B_{s, i}$ is only the
$i$th ballot drawn for seat $s$ and its sampling will take place at a
time $t \geqslant i$.

Let $n_s(t)$ denote the number of ballots drawn for seat $s$ up to and
inclusive of time $t$.
Define the process $M_{s, j} \equiv (M_{s, j, t})_{t \geqslant
0}$ by letting $M_{s, j, 0} \equiv 1$ and
\begin{equation} \label{assertion_level_mtg}
M_{s, j, t} \equiv
    \prod_{i = 1}^{n_s(t)} \left(1 + \lambda_{s, j, i}
                                          (X_{s, j, i} -
                                         \mu_{s, j, i})\right)
\quad \text{for all } t = 1, 2, \dots,
\end{equation}
where
$
\mu_{s, j, i} \equiv \frac{N_s/2 - \sum_{k = 1}^{i-1} X_{s, j, k}}{N_s - i + 1}
$
is the hypothesised mean of remaining assorter values just before the $i$th
ballot is drawn assuming $\bar{x}_{s,j} = 1/2$ (the boundary value for
$H_{s, j}$ to hold), and $\lambda_{s, j, i} \in (0, \mu_{s, j, i}^{-1})$
is a parameter depending only on $\cF_{T_{s}(i)-1}$, where
$T_s(i) \equiv \min \{t : n_s(t) = i\}$ is the first time that $i$ ballots
have been sampled for seat~$s$.  Each $\lambda_{s, j, i}$ controls how
aggressively the process responds to the value $X_{s, j, i} - \mu_{s, j, i}$,
and can be set based on the information available from all the revealed ballots
before the $i$th ballot is drawn for seat $s$. Larger values of $\lambda_{s, j,
i}$ produce faster growth of $M_{s, j, t}$ when $H_{s, j}$ is false.

Note that $M_{s, j}$ is an example of a \emph{betting process}
\citep{waudbysmith2024estimating}, introduced for election auditing by the
ALPHA process of \citet{stark2023alpha}.
Unlike the original ALPHA, $M_{s, j}$ has the variation that
its value may not update at every time step because sampling doesn't
necessarily occur for seat $s$ at all $t$.  Alternatively, we can write
\begin{equation} \label{assertion_level_mtg_alt_expr}
M_{s, j, t} = M_{s, j, t - 1} \cdot
    \left\{D_{s, t} \cdot
        \left(1 + \lambda_{s, j, n_s(t)}
                       (X_{s, j, n_s(t)} -
                      \mu_{s, j, n_s(t)})\right)  + 1 - D_{s, t}\right\}
\end{equation}
which gives $M_{s, j, t} = M_{s, j, t - 1}$ when $D_{s, t} =0$
and elucidates how the betting process evolves from $t - 1$ to $t$.

Importantly, $M_{s, j}$ is a \emph{test supermartingale} under $H_{s, j}$.
Specifically, it is non-negative
($M_{s, j, t} \geqslant 0$ for all $t \geqslant 0$),
has initial value 1 ($M_{s, j, 0} = 1$), and
\begin{equation}\label{super_mtg_property_Mcjt}
\bE [M_{s, j, t} \mid \cF_{t-1}] \leqslant M_{s, j, t - 1}
\end{equation}
when $H_{s, j}$ is true.  We establish these facts in
\appendixref{app:pf_non_neg_supermartingale}.
Ville's inequality~\citep{ville1939etude} then implies that $M_{s, j}$
is anytime-valid for testing $H_{s, j}$.

To test the seat as whole, we combine these together into a \emph{seat-level}
process
\begin{equation} \label{option_1_construction}
E_{s, t} \equiv \min_{1 \leqslant j \leqslant J_s} M_{s, j, t}.
\end{equation}
By virtue of the `min' construction, this is anytime-valid for testing $H_s$
(see \autoref{thm:main}, below).
However, our goal is not test each seat on its own, but to use these processes
as a building block for testing more complex hypotheses.

\subsection{A test process for certifying a parliamentary majority}
\label{sec:certify_parliament}

Let $\cC \subset \cW$ be a subset of the reported winning seats, and consider
\begin{equation} \label{H_cC_as_intersection}
H_{1/\cC} \equiv \bigcap_{s \in \cC} H_s,
\end{equation}
the \emph{intersection null hypothesis} that $H_s$ is true for all seats
in $\cC$.
Note that $H_{1/\cC}$ is true when \emph{no} ($<1$) seats in $\cC$ are truly
won by the winning party, and hence its notational similarity to $H_{r/\cW}$.
Now define
\begin{equation} \label{intersection_E_t}
E_{1/\cC, t} \equiv \prod_{s \in \cC} E_{s, t}.
\end{equation}
This provides a stepping-stone to a process for testing $H_{r/\cW}$, because we
can write that hypothesis as the following union
\begin{equation} \label{H_rW_as_union}
H_{r/\cW} = \bigcup_{\substack{\cC \subset \cW : \\ |\cC| = |\cW| - r + 1}}
            H_{1/\cC}.
\end{equation}
Our test process for a parliamentary majority can now be defined as follows:
\begin{equation} \label{e_process_r_out_of_W}
E_{r/\cW, t} \equiv  \prod_{s = 1}^{|\cW| - r + 1} E_{(s), t } \\
= \min \{E_{1/\cC, t} : |\cC| = |\cW| - r + 1\}
\end{equation}
where $E_{(1), t} \leqslant \cdots \leqslant E_{(|\cW|), t }$ are the
\emph{order statistics} of $\{E_{s, t}\}_{s \in \cW}$ at time $t$.

In other words, $E_{r/\cW, t}$ is the product of the smallest $|\cW| - r + 1$
values among $\{E_{s, t }\}_{s \in \cW}$, and the second equality in
\eqref{e_process_r_out_of_W} follows naturally. It is easy to check that all of
$E_{s, t}$, $E_{1/\cC, t}$ and $E_{r/\cW, t}$ are computable from $\cF_t$.  The
following theorem proven in \appendixref{app:pf_main_thm} states that they
indeed define processes that are anytime-valid for testing $H_s$, $H_{1/\cC}$
and $H_{r/\cW}$.

\begin{theorem}  \label{thm:main}
The processes
$E_s       \equiv (E_{s, t})_{    t \geqslant 0}$,
$E_{1/\cC} \equiv (E_{1/\cC, t})_{t \geqslant 0}$ and
$E_{r/\cW} \equiv (E_{r/\cW, t})_{t \geqslant 0}$ are
anytime-valid for testing $H_s$, $H_{1/\cC}$ and $H_{r/\cW}$, respectively.
\end{theorem}

Intuitively, using $E_{r/\cW}$ to test $H_{r/\cW}$ makes sense: In light of the
union representation \eqref{H_rW_as_union}, $H_{r/\cW}$ is false if and only if
all intersection nulls $H_{1/\cC}$ with $|\cC| = |\cW| - r +1$ are false; the
`min' representation in \eqref{e_process_r_out_of_W} says that $E_{r/\cW, t}$
is large enough to reject $H_{r/\cW}$ only when, for all $\cC$ with cardinality
$|\cW| - r + 1$, the value of $E_{1/\cC, t}$ is large enough to reject the
corresponding hypothesis $H_{1/\cC}$.

\subsection{Related works}

In the statistics literature, $H_{r/\cW}$ is a \emph{partial conjunction}
hypothesis, testing whether at least $r$ of the seat-level nulls
$\{H_s\}_{s \in \cW}$ are false. The notion is due to
\citet{benjamini2008screening} in the context of \emph{replicability analysis};
\citet{bogomolov2023replicability} provide a recent survey.
Our construction sits in the \emph{e-value} and \emph{e-process} framework
\citep{vovk2021evalues, ramdas2023game}: each of $E_{s,t}$, $E_{1/\cC,t}$ and
$E_{r/\cW,t}$ is an e-process for its null, and combining the seat-level ones
into $E_{r/\cW,t}$ is a sequential Benjamini--Heller-type construction
\citep{hoang2022combining, gablenz2025catch}. As shown in
\appendixref{app:pf_main_thm}, it stays valid even when sampling induces
cross-seat dependence, extending \citet{vovk2024merging}.

\citet[Section~4]{mohanty2019auditing} first proposed certifying the overall
winning party rather than every seat, for a more efficient audit, and sketched
a test of $H_{r/\cW}$, but did not implement nor evaluate it. Like us, they
built an anytime-valid $E_{s,t}$ for each $H_s$, but then transformed it into a
p-value $P_{s,t} \equiv 1/E_{s,t}$ and combined these across each subset $\cC
\subset \cW$ by Fisher's method \citep{Fisher1973}.
This is the stratified
union-intersection idea of SUITE \citep{ottoboni2018suite}, applied across
seats rather than across strata of a single contest.
Working directly with the e-processes instead allows us to use arbitrary
sampling schemes $D_{s,t}$ with easy theoretical justification, including the
adaptive schemes in \autoref{sec:sampling_schemes} which induce dependence
across $s$, and is uniformly more efficient than the p-value approach; see
\appendixref{app:mohanty}.

A final pair of RLA methods also test a union-of-intersections null, but within
a \emph{single} contest.
AWAIRE \citep{ek2023adaptively} audits a single instant-runoff election.
Its component tests share the same population of ballots, so they are not
conditionally independent and must be combined by \emph{averaging} rather than
as a product.  In our setting, since each seat is a separate contest, their
e-processes are conditionally independent and can be multiplied.
\citet{spertus2026stratified} stratify a single contest and, like us, combine
by products.  They pool evidence about one contest-wide margin across strata.
In contrast, we pool evidence about many separate contests, of which only $r$
need to be verified.

%% ================================================================
\section{Adaptive sampling schemes}
\label{sec:sampling_schemes}
%% ================================================================

\begin{figure}[t]
\centering
\includegraphics[width=\textwidth]{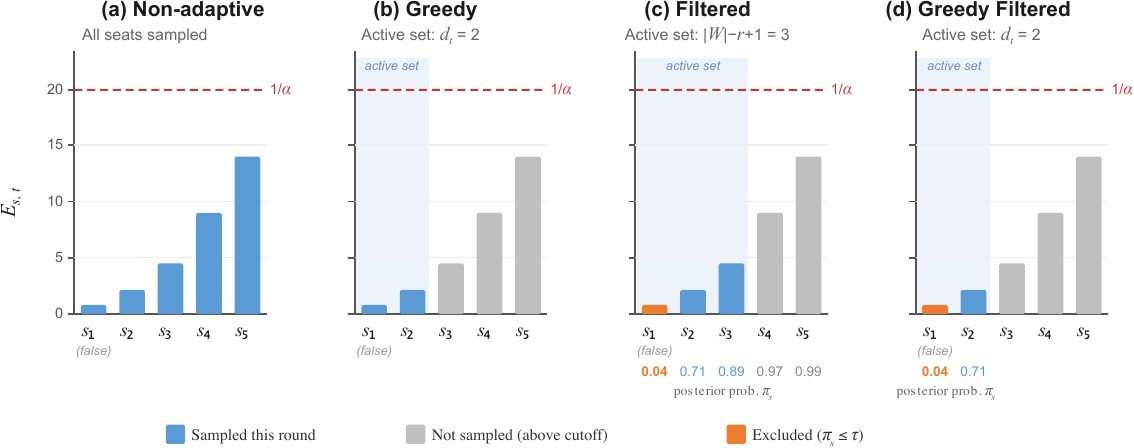}
\caption{Illustration of the four sampling schemes at a single round, with
$|\cW| = 5$ seats and $r = 3$.
Bars show the current values of $E_{s,t}$; the dashed red line is $1/\alpha$.
Blue bars indicate seats sampled in the next round; grey bars indicate seats
above the active-set cutoff; orange bars indicate seats excluded by the
posterior filter ($\pi_s \leqslant \tau$).
Seat~$s_1$ is falsely reported.
(a)~Non-adaptive: every seat is sampled.
(b)~Greedy: only the $d_t$ weakest seats are sampled.
(c)~Filtered: only seats among the $|\cW|-r+1$ weakest with $\pi_s > \tau$ are
    sampled.
(d)~Greedy Filtered: combines the $d_t$ active-set window of the Greedy scheme
    with the $\pi_s > \tau$ filter of the Filtered scheme.}
\label{fig:sampling_schemes}
\end{figure}

Given how $E_{r/\cW, t} $ is defined in \eqref{e_process_r_out_of_W}, the
ballot drawing decisions $D_{s, t}$ should ideally be made so that the product
of the smallest $|\cW| - r + 1$ values among $\{E_{s, t }\}_{s \in \cW}$ can
exceed $1/\alpha$ with minimal sampling.
We now describe several schemes for adaptively determining $D_{s, t}$.
These are illustrated in \autoref{fig:sampling_schemes}.

\subsubsection{Non-adaptive.} \label{sec:naive_scheme}
This scheme sets $D_{s, t} = 1$ for all $s \in \cW$ and all $t$. The auditor
makes no attempt to reduce the number of ballots to be sampled.

\subsubsection{Greedy.} \label{sec:greedy_scheme}
Since the winning party can be immediately certified at time $t$ as long as the
product of the smallest $|\cW| - r + 1$ values among $\{E_{s, t }\}_{s \in
\cW}$ exceeds $1/\alpha$, a natural heuristic is to `greedily' sample ballots
for the seats with the lowest seat-level processes, in the hope that the
additional ballots will increase these values.  Let $d_t \in \{1, \dots,
|\cW|\}$ be a cutoff sample size. The Greedy scheme sets
\[
D_{s,t+1} =
\begin{cases}
1, & \text{if } E_{s, t} \leqslant E_{(d_t), t}, \\
0, & \text{if } E_{s, t} > E_{(d_t), t},
\end{cases}
\]
where $E_{(1), t} \leqslant \cdots \leqslant E_{(|\cW|), t}$ are the order
statistics of $\{E_{s, t}\}_{s \in \cW}$ at time $t$.

The question is how to choose $d_t$. Define
\begin{equation} \label{eqn:delta_t}
\delta_t = \max\Big\{i : i \in \{1, \dots, r\} \text{ and }
\prod_{s = i}^{|\cW| - r + i} E_{(s), t} < 1/\alpha\Big\},
\end{equation}
with the convention that $\delta_t = 0$ if the maximum is taken over an empty
set (meaning the winning party can already be certified at time $t$). Let
\begin{equation} \label{eqn:d_t_choice}
d_t =
\begin{cases}
|\cW|,        & \text{if } \delta_t = r, \\
\delta_t + a, & \text{if } \delta_t < r,
\end{cases}
\end{equation}
for a constant $a \in \{0, 1, \dots, |\cW| - r\}$. When $\delta_t = r$, even
the product of the largest $|\cW| - r + 1$ elements among $\{E_{s, t}\}_{s \in
\cW}$ is below $1/\alpha$, so we sample all $|\cW|$ seats. When $\delta_t < r$,
the order statistics $E_{(1), t}, \dots, E_{(\delta_t), t}$ are precisely those
for which the product of any consecutive set of $|\cW| - r + 1$ containing them
would not exceed $1/\alpha$, and at a minimum we should gather more evidence
for the $\delta_t$ seats corresponding to these processes. The parameter $a$
controls how aggressively the scheme minimises per-round sampling: $a = 0$ is
the most aggressive, while larger values of $a$ hedge by sampling additional
seats beyond the strict minimum.

Since sampling is without replacement, it is possible for every one of the
$d_t$ selected weakest seats to have already exhausted its ballot supply (i.e.,
all $N_s$ ballots have been drawn) without achieving certification ($E_{r/\cW,
t} \geqslant 1/\alpha$).
In this case, the round would issue no new samples and the audit would stall.
To avoid this, we apply the following fallback: if all $d_t$ weakest seats are
exhausted, we incrementally increase $d_t$ by one---extending the selection to
the next-weakest seat---and repeat until either at least one selected seat has
remaining ballots or $d_t = |\cW|$.  This guarantees that in each round the
sampling effort is directed at the weakest seats whose ballot supply has not
yet been exhausted.

\subsubsection{Filtered.} \label{sec:bayesian_scheme}
The Greedy scheme concentrates all sampling effort on the seats with
the weakest seat-level processes. While intuitive, this can fail when some of
the weakest seats correspond to \emph{falsely reported} outcomes, i.e., seats
where the winning party did not actually prevail. For such a seat~$s$, the
process $E_s$ won't grow, which will lead to $E_{r/\cW}$ stalling or
growing very slowly.

We define a Bayesian inference approach that aims to filter out such seats.
For each seat $s \in \cW$, we monitor the observed ballots and update a
posterior distribution on the true ballot-type proportions.  After each round,
we use this to calculate a posterior probability, $\pi_s$, that we use to
exclude unpromising seats from further sampling.

Our approach uses a canonical Dirichlet--multinomial conjugate setup.  We omit
most of the details (they are standard) and just sketch out the main steps.

For seat~$s$ with $L_s$ ballot types, set a
prior on the ballot-type probabilities $\mathbf{p}_{s} = (p_{s,1}, \dots,
p_{s,L_s}) \sim \mathrm{Dirichlet}(\boldsymbol{\alpha}_{s,0})$, with
hyperparameters
$\boldsymbol{\alpha}_{s,0} \equiv \hat{\mathbf{p}}_{s,0} \, \kappa_{s, 0}$
based on an initial value $\hat{\mathbf{p}}_{s,0}$ and concentration parameter
$\kappa_{s,0}$.
After observing $n_s(t)$ ballots at time $t$, use conjugacy\footnote{%
The conjugacy relies on the sampled ballots being treated as multinomial, an
approximation since sampling is without replacement. This model is used only to
choose where the filter samples next, which is a predictable decision, so by
\autoref{thm:main} the risk limit holds regardless of how accurate the model
is.}
to give the posterior $\mathbf{p}_s \mid \cF_t \sim
\mathrm{Dirichlet}(\boldsymbol{\alpha}_{s,n_s(t)})$.
Let $\theta_{s,j}$ be the true assorter mean for assertion $j$.
The posterior on $\mathbf{p}_s$ induces a posterior on $\theta_{s,j}$ via its
assorter function.
The joint posterior on $\theta_{s,j}$ across all $j$ could be used to calculate
the posterior probability that seat~$s$ was truly won ($\theta_{s,j} > 1/2$ for
all $j$).  For convenience, and for a fixed $\epsilon \geqslant 0$, we instead
define
\begin{equation} \label{eqn:post_prob_general}
\pi_s \equiv \min_{1 \leqslant j \leqslant J_s}
    \Pr(\theta_{s,j} > 1/2 + \epsilon \mid \cF_t),
\end{equation}
so that $\pi_s$ is small when at least one assertion has substantial posterior
evidence that its margin does not exceed $\epsilon$.
We investigate only
$\epsilon = 0$, which deprioritises seats that appear falsely reported. A
positive $\epsilon$ would additionally deprioritise seats won by too small a
margin to certify feasibly. (Determining an efficient, adaptive choice of
$\epsilon$ is left for future work.)
We calculate each $\Pr(\theta_{s,j} > 1/2 + \epsilon \mid \cF_t)$ via a normal
approximation to the posterior, or an exact calculation in the case of
$L_s = 2$ (two ballot types).

The Filtered scheme restricts attention to the $|\cW| - r + 1$ seats with the
smallest current seat-level processes $E_{s, t}$ since only these seats enter
the product defining $E_{r/\cW, t}$ in \eqref{e_process_r_out_of_W}.  Among
this active set, it samples those whose posterior probability exceeds a
threshold $\tau$:
\begin{equation} \label{eqn:bayesian_lambda}
D_{s,t+1} =
\begin{cases}
1, & \text{if } E_{s, t} \leqslant E_{(|\cW|-r+1), t} \text{ and }
                \pi_s > \tau, \\[4pt]
0, & \text{otherwise}.
\end{cases}
\end{equation}
Choosing $\tau$ small (we set $\tau = 0.01$) makes the filter liberal: a seat
is excluded only when there is substantial evidence that it was \emph{not}
truly won.
If no seat in the active set satisfies both $\pi_s > \tau$ and has remaining
ballots, the $\pi_s > \tau$ condition is dropped and $D_{s, t+1} = 1$ is set
for every seat in the active set that still has ballots to draw.

\subsubsection{Greedy Filtered.} \label{sec:greedy_bayesian_scheme}
The Greedy scheme is efficient in benign cases (no falsely reported seats),
while the Filtered scheme is more robust when the audit encounters false seats.
We combine them together by using both the adaptive window $d_t$ (the `active
set') and the $\pi_s > \tau$ selection filter.

To ensure we always draw at least one ballot, we use the following fallback
rules:
(1)~The active set is expanded one seat at a time---adding the next-weakest
seat---first to ensure it contains a seat with remaining ballots, and, if
necessary, further (up to a maximum size of $|\cW| - r + 1$) to ensure it
contains a seat that also satisfies $\pi_s > \tau$.
(2)~Similar to the fallback of the Filter method, if the above rule selects no
seats, the $\pi_s > \tau$ filter is dropped and the scheme sets
$D_{s, t+1} = 1$ for every seat in that window with remaining ballots.

%% ================================================================
\section{Results}
\label{sec:results}
%% ================================================================

We evaluate the sampling schemes on plurality elections audited by
ballot-polling\footnote{%
Ballot-polling audits require only the paper trail and the reported outcome, so
can be used for any election with a paper trail.  This is the simplest scenario
and the only one we investigate here.
More efficient audits are possible if electronic records of the ballots are
available; only some jurisdictions would have these.%
}
without replacement.
Throughout, we assume there are no invalid ballots\footnote{%
This is purely for simplicity. The methods can cope with invalid ballots: the
assorters would score these as $1/2$ (see \autoref{sec:testing_one_seat}).}
and set a risk limit of $\alpha = 0.05$.
\autoref{sec:sim_plurality} shows results from controlled two-candidate
simulations (with further results in \appendixref{app:further_simulations}),
while in \autoref{sec:sim_india} we simulate audits based on the
2014 Indian parliamentary election. Code and data to reproduce our results
are available at:
\url{https://github.com/freejstone/RLA_parliamentary_majorities_code}.

\subsubsection{Audit strategies.}
We compare six strategies for certifying the parliamentary majority (testing
$H_{r/\cW}$):
Non-adaptive;
Greedy with $a = 0$ or $a = 3$ in \eqref{eqn:d_t_choice};
Filtered with $\tau = 0.01$;
and Greedy Filtered with $a = 0$ or $a = 3$, and $\tau = 0.01$.

For comparison, we include two further strategies to act as benchmarks:
(1)~\emph{Top-$r$ seats} is a two-phase rule that first samples from $\cR$,
the \(r\) winning seats with the largest reported seat margins, switching to
the remaining $|\cW| - r$ winning seats if the first $r$ seats are exhausted.
During the first phase, it effectively attempts to certify all seats in $\cR$,
i.e., to reject $H_{r/\cR}$.
(2)~\emph{All seats} aims to certify all reported winning seats, rather than
just a majority. It uses the same sampling rule as \textit{Non-adaptive}, but
sets $r = |\cW|$, giving it the stricter target of rejecting
$H_{|\cW|/\cW}$.

\subsubsection{Tuning parameters.}

For testing each assertion, we use $M_{s,j}$
from~\eqref{assertion_level_mtg} and the truncated-shrinkage strategy of
ALPHA~\citep{stark2023alpha} to set $\lambda_{s,j,i}$.
This uses tuning parameters $d$ and $\eta_0$, which we set to
$d = 200$ and $\eta_0 = 0.51$ as per \citet{ek2024efficient}.

For the Dirichlet prior used in the filtering strategies, we set the
hyperparameters such that the prior mean for each assertion was 0.51 and
$\kappa_{s,0} = 200$ (for plurality voting this involves solving a set
of linear equations).  This was to mimic the role of $\eta_0$ and $d$
in the ALPHA truncated-shrinkage estimator.

\subsection{Simulated plurality contests} \label{sec:sim_plurality}

We first evaluated the audit strategies using synthetic two-candidate plurality
contests. Consider a parliament with \(|\cS|=100\) seats, so a majority
requires $r = \lfloor |\cS|/2 \rfloor + 1 = 51$ seats. We fixed the number of
reported winning seats at $|\cW| = 60$ and each seat to have $N_s = 5000$
ballots.

For each seat $s$, we set a target proportion of ballots for the reported
winner, $p_{\mathrm{target},s}$.  The reported winning candidate receives
$\lfloor N_s \times p_{\mathrm{target},s}\rfloor$ ballots and the remaining
ballots are assigned to the opposing candidate.

\begin{figure}[tb]
\centering
\includegraphics[width=\textwidth]{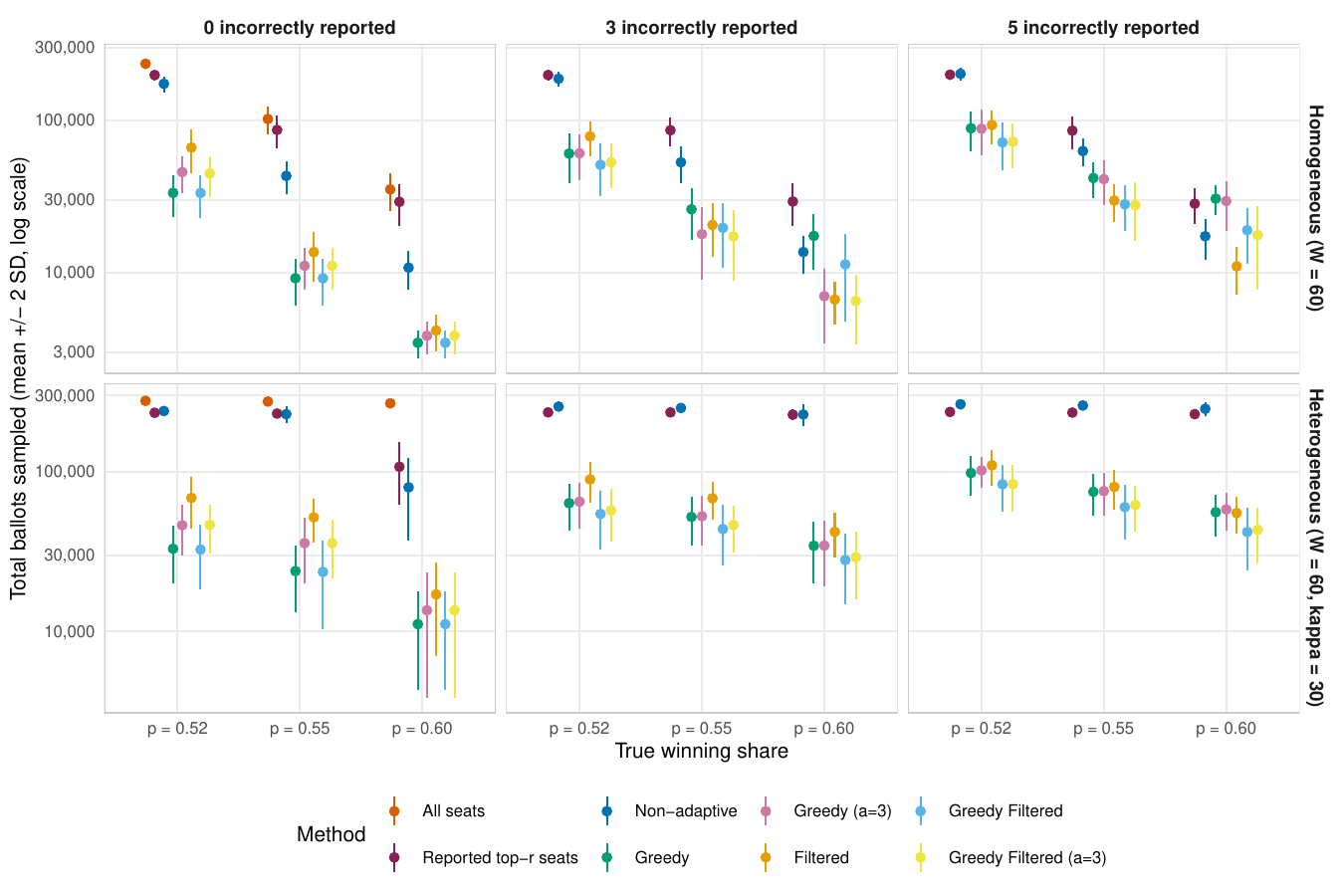}
\caption{Simulated two-candidate plurality contests. The total
number of ballots sampled to certification: each point shows the mean
across 100 replicates and error bars denote $\pm 2$ standard deviations.
Columns correspond to the number of falsely reported seats, $n_\text{false}$.
Rows correspond to different ways to set the margins:
top row shows Scenario~1,
bottom row shows Scenario~2.
Results for the \emph{All seats} method are only reported for
$n_{\mathrm{false}} = 0$, since
for $n_{\mathrm{false}} > 0$ it tests $H_{|\cW|/\cW}$, which is a true
null and requires a full recount with probability at least $1 - \alpha$.}
\label{fig:sim1_sim2}
\end{figure}

We allowed some reported winning seats to be false. If
$n_{\mathrm{false}} > 0$, then $n_{\mathrm{false}}$ of the seats in $\cW$
were generated with $p_{\mathrm{target},s} = 0.48$,
so that the reported winner did not truly win that seat.
The remaining $|\cW| - n_{\mathrm{false}}$ seats are correctly reported;
we considered two scenarios for setting $p_{\mathrm{target},s}$ for these
seats:
\begin{enumerate}
\item (Homogeneous margins)
All truly won seats share the same target vote share
$p_{\mathrm{target},s} = p_{\mathrm{target}}$.
We varied
$p_{\mathrm{target}}\in\{0.52,0.55,0.60\}$.

\item (Heterogeneous margins)
The target vote share for each truly won seat was drawn independently as
\[
p_{\mathrm{target},s}
\sim
\mathrm{Beta}\!\left(
\bar p_{\mathrm{target}}\kappa_{\mathrm{het}},
(1-\bar p_{\mathrm{target}})\kappa_{\mathrm{het}}
\right),
\]
and then truncated to lie in $[0.51, 0.999]$, ensuring that the reported
winner truly wins the seat. We fixed $\kappa_{\mathrm{het}} = 30$,
corresponding to moderate heterogeneity, and varied
$\bar p_{\mathrm{target}} \in \{0.52,0.55,0.60\}$.
\end{enumerate}
We also varied $n_{\mathrm{false}}\in\{0,3,5\}$, giving
$3 \times 3 = 9$ configurations for each of the two scenarios.
We replicated each one 100 times.
\autoref{fig:sim1_sim2} shows our results, in terms of the number of ballots
sampled until certification.

All six strategies for certifying the parliamentary majority were
markedly more efficient than the two benchmarks.
\emph{All seats} necessarily uses the most ballots: certifying $H_{|\cW|/\cW}$
requires \emph{every} reported seat to individually exceed the threshold.
\emph{Top-$r$ seats} is handicapped in the same way: during its first phase it
effectively certifies $H_{r/\cR}$, which is again driven by the minimum
process.
In contrast, certifying $H_{r/\cW}$ depends on the product of the
bottom $|\cW| - r + 1$ seat-level processes, so stronger seats can carry the
minimum.
Among the majority schemes, \emph{Non-adaptive} was least efficient, since it
samples every seat regardless of how informative each is.
The `filtered' variants outperform their pure-greedy counterparts once
$n_{\mathrm{false}} > 0$, as expected, by redirecting effort away from
seemingly false seats.

We provide some further results in \appendixref{app:further_simulations}, in
which we also vary the number of reported winning seats
$|\cW| \in \{51, 60, 80\}$ and the heterogeneity parameter
$\kappa_{\text{het}} \in \{10, 30, 100\}$.
Our broad conclusions remained the same.

\subsection{Indian general election} \label{sec:sim_india}

To evaluate the audit framework on a realistic parliament-scale election, we
simulated audits using data from the 2014 Indian general election of
the Lok Sabha (Lower House). We used candidate-level vote totals for all 543
constituencies from a community-scraped ECI dataset\footnote{
\url{https://github.com/datameet/india-election-data/tree/master/parliament-elections/election2014},
accessed on 2026-04-19.},
originally sourced from the Election Commission of India's results portal. The
Bharatiya Janata Party (BJP) won $|\cW| = 282$ of the 543 seats, forming a
single-party majority.
Certifying a BJP majority requires verifying that at least
$r = \lfloor 543 / 2 \rfloor + 1 = 272$
of these 282 reported winning seats were truly won.

Across the 282 BJP seats, the number of candidates ranged from $L_s = 5$ to
$L_s = 43$ (median 15), and the number of ballots per seat ranged from
approx.\ 87,000 to 1,512,000 (median 1,010,000), summing to a total of about
282~million ballots.
The winner's vote share ranged from 0.264 to 0.758 (median 0.495).
Because each $L_s$-candidate plurality contest has
$J_s = L_s - 1$ SHANGRLA assertions, the minimum assorter mean across these
head-to-head comparisons is the bottleneck for certifying a seat. In the BJP
seats, this minimum assorter mean ranged from 0.5002 to 0.7812 (median 0.5831).

We generated $n_{\mathrm{false}}$ falsely reported seats by swapping the vote
totals of the reported winner and the runner-up.
The BJP remains the reported winner, but the runner-up is the true winner in
the underlying ballot population. We apply
this perturbation to the $n_{\mathrm{false}}$ most marginal BJP seats, where
marginality is measured by the two-candidate margin between the BJP candidate
and the runner-up. We vary $n_{\mathrm{false}} \in \{0,3,5\}$.
We replicated each of these three configurations 100 times.

\begin{figure}[tb]
\centering
\includegraphics[width=\textwidth]{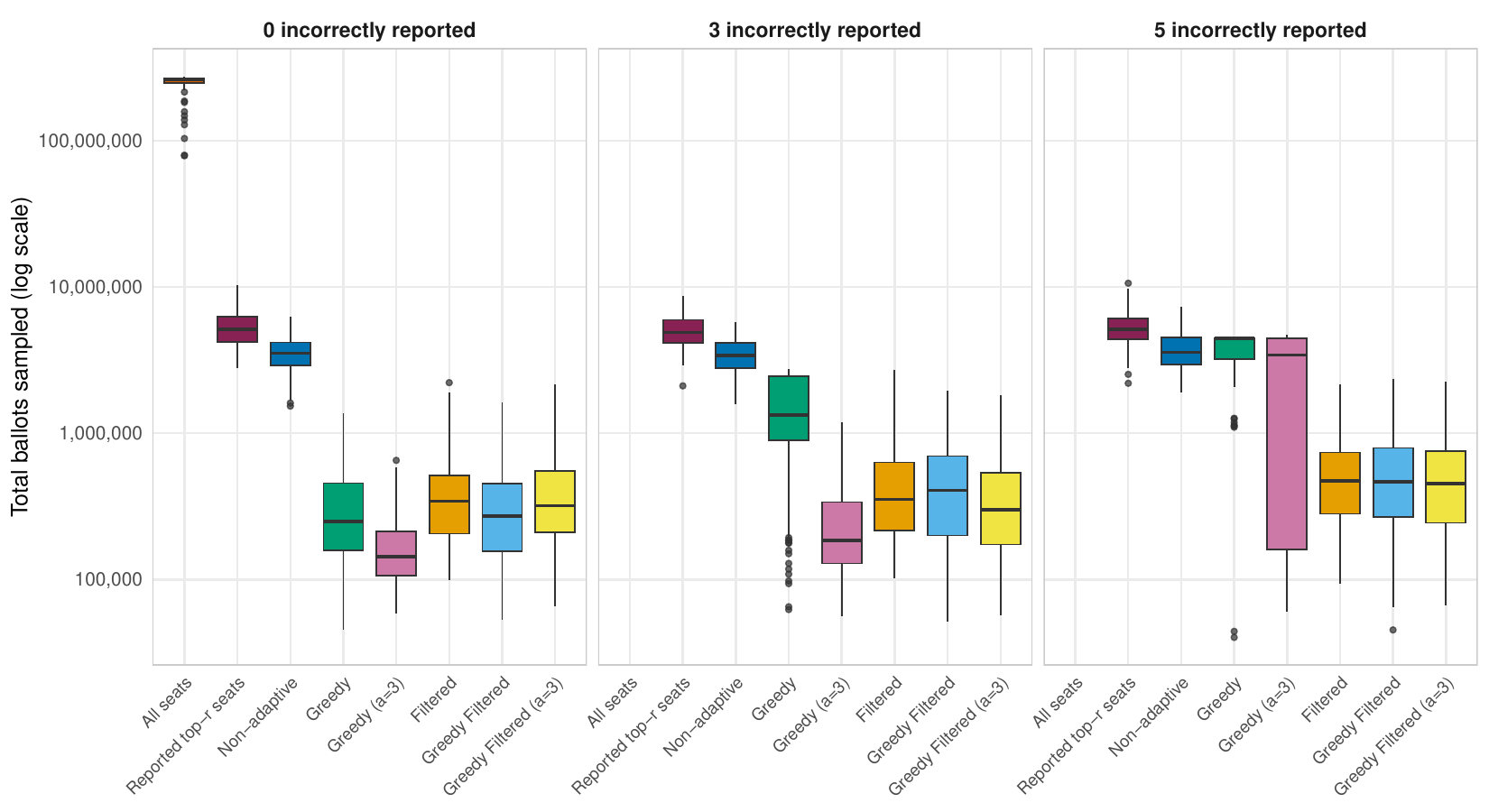}
\caption{India 2014. Box plots of the total number of ballots sampled to
    certification across 100 replicates, with one panel per number of
    falsely reported seats~$n_{\mathrm{false}}$.
    Results for the \emph{All seats} method are omitted for
    $n_{\mathrm{false}} > 0$, like in \autoref{fig:sim1_sim2}.}
\label{fig:india_results}
\end{figure}

\autoref{fig:india_results} reports the total number of ballots sampled before
certification.
We see qualitative behaviour similar to \autoref{sec:sim_plurality}.
All adaptive schemes dominated the benchmarks by one to three orders of
magnitude: the \emph{All seats} baseline required roughly 250~million
ballots, and the \emph{Reported top-$r$ seats}
benchmark required around 5~million across all $n_{\mathrm{false}}$, whereas
the others typically required only a few hundred thousand.
Within the parliamentary-majority schemes, the \emph{Greedy} ($a=0$) scheme
degraded sharply as we increased $n_{\mathrm{false}}$ (going from about 400k to
3.8M ballots),
demonstrating the price of pouring effort into false seats.
Widening the active set to $a=3$ restored robustness at $n_{\mathrm{false}}=0$,
where it is the most efficient scheme overall.
However, once $n_{\mathrm{false}} > 0$, the `filtered' schemes took over,
showing consistently superior performance when $n_{\mathrm{false}} = 5$.

%% ================================================================
\section{Discussion}
%% ================================================================

We have developed a risk-limiting audit for parliamentary majorities.  We
formalised this task as a partial conjunction hypothesis test and constructed a
sequential parliament-level audit statistic by combining seat-level statistics.
The resulting procedure controls the risk of certifying an incorrect majority
outcome, while allowing the auditor to adaptively allocate sampling effort
across seats.

Although our experiments focused on ballot-polling audits of plurality
elections, the methodology is not limited to this setting. Because it is built
on SHANGRLA assertions, it inherits that framework's generality, covering a
wide variety of social choice functions, other audit styles such as comparison
audits, and settings with missing ballots. It can stretch further still,
targeting \emph{any} hypothesis of the form $H_{r/\cW}$ for any value of $r$,
not just majorities.

In practice, each constituency needs the same infrastructure as any
single-contest RLA.
The extra burden imposed by the majority audit is cross-seat coordination:
a central authority must aggregate data from the sampled ballots, update
$E_{r/\cW,t}$, and issue the next round's instructions.
This overhead favours a few large-batch rounds rather than the single-ballot
rounds we used for exposition (n.b., the risk limit holds for any batch size).
Designing this coordination to run efficiently in practice is an important
problem in its own right, one we leave for future work. The burden is
nonetheless far lighter than auditing every seat: the total sample is orders of
magnitude smaller, and the adaptive schemes concentrate effort on a few seats,
allowing most constituencies to stop early.

\enlargethispage{\baselineskip}

In our experiments we only considered ballot-polling audits.  When the relevant
election infrastructure is available, more sophisticated and efficient audit
designs are possible, such as ballot-level comparison audits.  The sample size
required for such audits typically grows linearly with respect to the
reciprocal of the margin, whereas for ballot-polling it grows quadratically
\citep{lindeman2012gentle}.  Thus, combining our method with a comparison audit
should lead to even smaller sample sizes, in the order of thousands of ballots
for the Indian election example.

Our audit certifies that a reported winning party won a majority of seats in
its own right. This is the pertinent question when a single party, or a
pre-declared alliance, wins an outright majority, as in our Indian example.
When no party/alliance wins such a declared majority, our method is \emph{not}
relevant: if a governing coalition is negotiated only after the election, the
majority to certify is not known while the ballots remain available, and such
elections are better served by auditing the individual seats.  However, because
our audit is built from ordinary seat-level assertions, it can be run in tandem
with a full seat-by-seat audit: the majority can be certified `midstream', as
soon as the parliament-level statistic $E_{r/\cW,t}$ crosses $1/\alpha$, while
the seat-by-seat audit continues towards certifying every individual seat or
recounting those that appear miscounted.

Several directions remain for future work:
(1)~It would be useful to compare our sampling strategies in diverse scenarios,
such as other election and audit types.
(2)~The efficiency of the audit depends on the choice of sampling scheme and
ALPHA tuning parameters; future work could study alternative choices.
This includes investigating the benefit of setting them based on reported vote
totals/margins, in scenarios where these are available.
(3)~Determining optimal ways to set sample sizes per seat for each round would
be useful for practice.
(4)~The process $E_{r/\cW}$ is anytime-valid for testing $H_{r/\cW}$, but
might not be optimal. Similarly, the process $E_{s,t}$ is one option for
summarizing seat-level information, but might not be optimal.
Investigating if there are more efficient alternatives would be of interest.

%% ================================================================

\bibliographystyle{splncsnat}
\bibliography{parliament_bib}

%% ================================================================

\appendix

%% ================================================================
\section{Proof that $M_{s, j}$ is a test supermartingale under $H_{s, j}$}
\label{app:pf_non_neg_supermartingale}
%% ================================================================

Non-negativity holds since $X_{s, j, i} \geqslant 0$ and $\lambda_{s, j, i}
\leqslant \mu_{s, j, i}^{-1}$ in \eqref{assertion_level_mtg}, and
$M_{s, j, 0} = 1$ by definition.
To prove \eqref{super_mtg_property_Mcjt}, from
\eqref{assertion_level_mtg_alt_expr} we can see that, when $H_{s,j}$ is true,
\begin{align*}
&\bE [M_{s, j, t} \mid \cF_{t -1}] \\
&= M_{s, j, t - 1} \left\{D_{s, t} \cdot \left(1 + \lambda_{s, j, n_s(t)}
    (\bE[X_{s, j, n_s(t)} \mid \cF_{t-1}] - \mu_{s, j, n_s(t)})\right) +
    1 - D_{s, t}\right\} \\
& \leqslant M_{s, j, t - 1},
\end{align*}
where the equality holds because the quantities $M_{s, j, t-1}$, $D_{s, t}$,
$\lambda_{s, j, n_s(t)}$ and $\mu_{s, j, n_s(t)}$ only depend on
$\cF_{t-1}$,\footnote{%
Here $\lambda_{s, j, n_s(t)}$ depends on $\cF_{T_s(n_s(t)) - 1}$, and
$\cF_{T_s(n_s(t)) - 1} \subset \cF_{t-1}$ since $T_s(n_s(t)) \leqslant t$.
Likewise $\mu_{s, j, n_s(t)}$ depends only on the first $n_s(t) - 1$ ballots
for seat $s$, all revealed by time $t-1$.}
and the inequality holds because $\bE[X_{s, j, n_s(t)} \mid
\cF_{t-1}] \leqslant \mu_{s, j, n_s(t)}$ under $H_{s, j}$: here $\mu_{s, j,
n_s(t)}$ is the mean of the remaining assorter values under $\bar{x}_{s,j} =
1/2$, the boundary (hence maximal) value for $H_{s, j}$.

%% ================================================================
\section{Proof of \autoref{thm:main}}
\label{app:pf_main_thm}
%% ================================================================

\subsubsection{$E_s$ is anytime-valid for $H_s$:}
Suppose $H_s$ is true. By the union \eqref{union_null}, some $j_s \in \{1,
\dots, J_s\}$ has $H_{s, j_s}$ true, and \eqref{option_1_construction} gives
$E_{s, t} \leqslant M_{s, j_s, t}$ for all $t \geqslant 0$. It thus suffices to
show that $\Pr(\sup_{t \geqslant 0} M_{s, j_s, t} \geqslant 1/\alpha) \leqslant
\alpha$ under $H_{s, j_s}$. This holds because $M_{s, j_s}$ is a test
supermartingale (\appendixref{app:pf_non_neg_supermartingale}) and Ville's
inequality \citep{ville1939etude}.

\subsubsection{$E_{1/\cC}$ is anytime-valid for $H_{1/\cC}$:}
Suppose $H_{1/\cC} = \bigcap_{s \in \cC} H_s$ is true, so $H_s$ holds for all
$s \in \cC$. By the union \eqref{union_null}, each $s \in \cC$ has an assertion
$j_s \in \{1, \dots, J_s\}$ with $H_{s , j_s}$ true; use these to define
$M_{1/\cC, t} \equiv \prod_{s \in \cC} M_{s, j_s, t}$. From
\eqref{option_1_construction}, $E_{1/\cC , t} \leqslant M_{1/\cC , t}$ for all
$t \geqslant 0$.
By this bound, it suffices to show that $M_{1/\cC}$ is a test supermartingale
and apply Ville's inequality, as follows.
From the expression in \eqref{assertion_level_mtg_alt_expr}, we have that
\begin{align*}
&\bE[M_{1/\cC, t} \mid \cF_{t-1}] \\
&= M_{1/\cC, t-1} \,
    \bE\left[\prod_{s \in \cC} \left\{D_{s, t} \cdot
    \left(1 + \lambda_{s, j_s, n_s(t)} (X_{s, j_s, n_s(t)} -
                                      \mu_{s, j_s, n_s(t)})\right)
    + 1 - D_{s, t}\right\}
    \mid \cF_{t-1}\right] \\
&\overset{\tiny{(a)}}{=}
    M_{1/\cC, t-1}
    \prod_{s \in \cC } \left\{D_{s, t} \cdot \left(1 + \lambda_{s, j_s, n_s(t)}
    (\bE[X_{s, j_s, n_s(t)} \mid \cF_{t-1}] -
       \mu_{s, j_s, n_s(t)})\right) +
    1 - D_{s, t}\right\} \\
&\overset{\tiny{(b)}}{\leqslant}
    M_{1/\cC, t-1},
\end{align*}
where $(a)$ holds because, conditional on $\cF_{t-1}$, the factors $D_{s,t}$,
$\lambda_{s, j_s, n_s(t)}$ and $\mu_{s, j_s, n_s(t)}$ are constants and the
$X_{s, j_s, n_s(t)}$ are independent across $s \in \cC$; $(b)$ follows as in
\appendixref{app:pf_non_neg_supermartingale}.

Since $M_{s, j_s, t} \geqslant 0$ and
$M_{1/\cC, 0} = \prod_{s \in \cC} M_{s, j_s, 0} = 1$, $M_{1/\cC}$ is a test
supermartingale, and Ville's inequality gives anytime-validity for $H_{1/\cC}$.

\subsubsection{$E_{r/\cW}$ is anytime-valid for $H_{r/\cW}$:}
Suppose $H_{r/\cW}$ is true. By \eqref{H_rW_as_union}, there is a subset
$\cC_0 \subset \cW$ with $|\cC_0| = |\cW| - r + 1$ and $H_{1/\cC_0}$ true. From
above, $E_{1/\cC_0}$ is anytime-valid for testing $H_{1/\cC_0}$, so
$\Pr(\sup_{t} E_{1/\cC_0, t} \geqslant 1/\alpha) \leqslant \alpha$. The `min'
representation \eqref{e_process_r_out_of_W} gives
$E_{r/\cW, t} \leqslant E_{1/\cC_0, t}$ for all $t$, so
$\Pr(\sup_{t} E_{r/\cW, t} \geqslant 1/\alpha) \leqslant
\Pr(\sup_{t} E_{1/\cC_0, t} \geqslant 1/\alpha) \leqslant \alpha$, proving the
claim.

%% ================================================================
\section{Comparison with the Fisher combination approach of
\citet{mohanty2019auditing}}
\label{app:mohanty}
%% ================================================================

We first show that our direct way of combining the seat-level
e-processes via products is substantially more efficient than the Fisher
combination proposal of \citet[Section~4]{mohanty2019auditing}, holding the
sampling scheme fixed and common to both methods.

Fix a time $t$ and let $k \equiv |\cW| - r + 1$. Certifying $H_{r/\cW}$
requires rejecting the intersection null $H_{1/\cC}$ for every $\cC \subset
\cW$ with $|\cC| = k$ \eqref{H_rW_as_union}. For a given $\cC$, our process
uses $E_{1/\cC, t} = \prod_{s \in \cC} E_{s, t}$ and rejects when it reaches
$1/\alpha$ (\autoref{thm:main}). \citet{mohanty2019auditing} instead form
seat-level p-values $P_{s, t} \equiv 1/E_{s, t}$ and combine them by Fisher's
method \citep{Fisher1973} to form the statistic
\[
-2 \sum_{s \in \cC} \ln P_{s, t} = 2 \ln \prod_{s \in \cC} E_{s, t}.
\]
This Fisher combined statistic is referenced to a standard $\chi^2_{2k}$
distribution under the null $H_{1/\cC}$. That is, their test rejects
$H_{1/\cC}$ when $\prod_{s \in \cC} E_{s, t} \geqslant \exp(\chi^2_{2k,
1-\alpha} / 2)$, where $\chi^2_{2k, 1-\alpha}$ is the $1-\alpha$ quantile of
the $\chi^2_{2k}$ distribution. In both cases, the critical subset $\cC$ is the
one with the smallest product, namely the one with the $k$ smallest seat-level
processes, so the majority decision reduces to a threshold on $\prod_{s=1}^{k}
E_{(s), t}$. Our method uses $1/\alpha$, theirs uses $\exp(\chi^2_{2k,
1-\alpha}/2)$.

The two thresholds coincide only when $k = 1$, since $\chi^2_{2, 1-\alpha} =
2\ln(1/\alpha)$. For $k \geqslant 2$, the Fisher threshold is strictly larger,
so our test rejects whenever theirs does and is uniformly at least as powerful.
At $\alpha = 0.05$, our threshold is $1/\alpha = 20$, while theirs is
${\approx}115$ at $k = 2$ and ${\approx}2.3 \times 10^{7}$ at $k = 11$ (ten
spare seats).  Because a seat-level process grows geometrically in the number
of ballots drawn, this gap inflates the required sample by a factor of about
$\chi^2_{2k, 1-\alpha}/(2\ln(1/\alpha))$, roughly $5.7$ at $k = 11$, which is a
substantial difference in efficiency.

Finally, we point out that \citet{mohanty2019auditing} justified the
theoretical validity of their Fisher combination proposal incorrectly. They
mistakenly treated their p-value processes $P_{s, t}$ as independent across
seats, and hence the typical quantile $\chi^2_{2k, 1-\alpha}$ under Fisher's
combining method can be used as the rejection threshold. However, this proof
argument only works under the non-adaptive sampling scheme in
\autoref{sec:naive_scheme}, because all the other sampling schemes in
\autoref{sec:sampling_schemes} necessarily inject dependence between the
e-processes (and therefore the transformed p-value processes) across seats. In
contrast, our exposition above actually proves the anytime validity of their
method correctly, by demonstrating that theirs is simply uniformly more
conservative than our direct e-process-based approach.

%% ================================================================
\section{Further simulations}
\label{app:further_simulations}
%% ================================================================

\begin{figure}
\centering
\includegraphics[width=\textwidth]{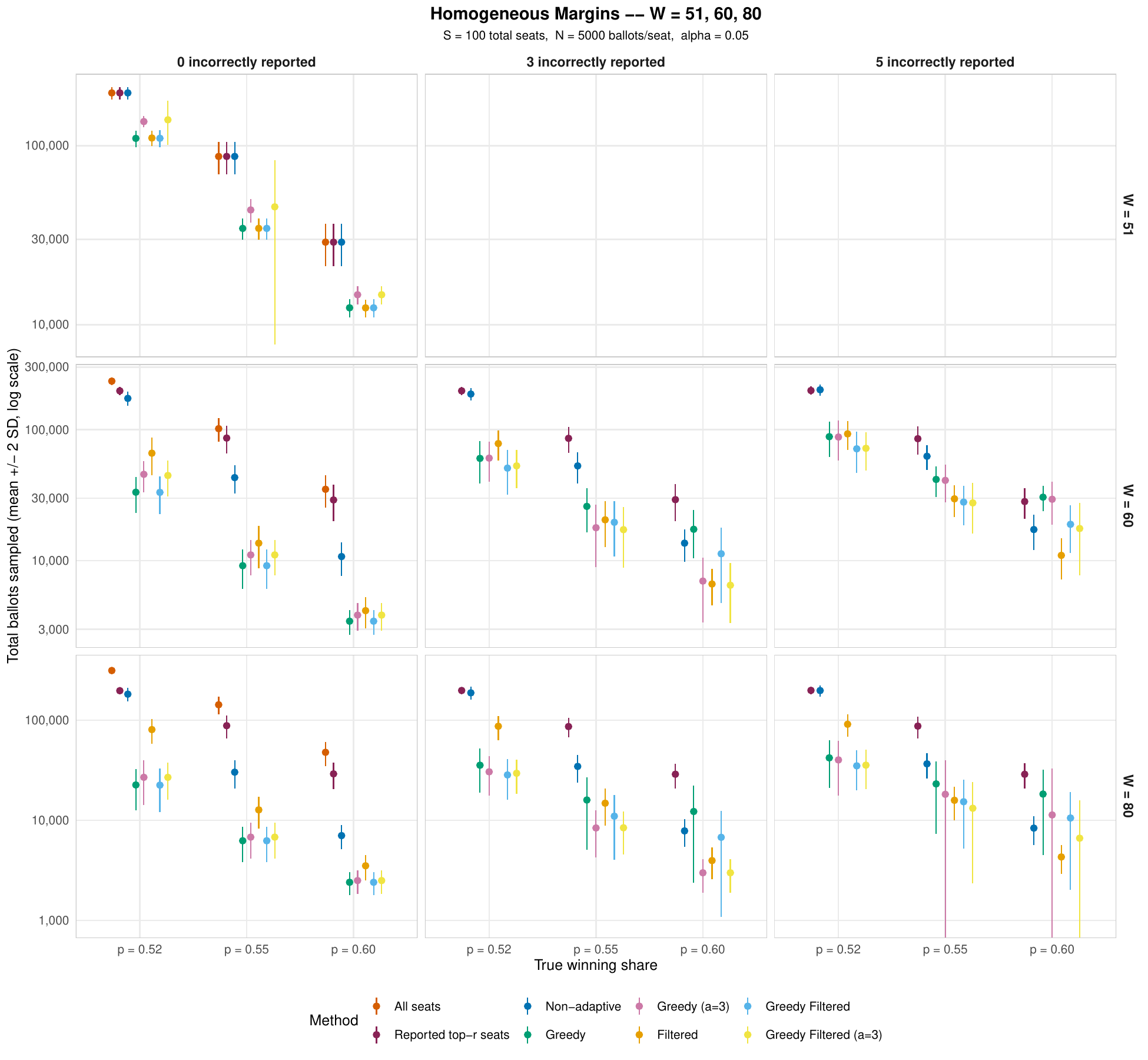}
\caption{Extended simulation results for the two-candidate plurality contests
of Scenario~1 from \autoref{sec:sim_plurality}.
The plot matches the first row of \autoref{fig:sim1_sim2}, but additionally
varies $|\cW| \in \{51, 60, 80\}$.}
\label{fig:sim1_extended}
\end{figure}

\begin{figure}[p]
\centering
\includegraphics[page=1,width=\textwidth]{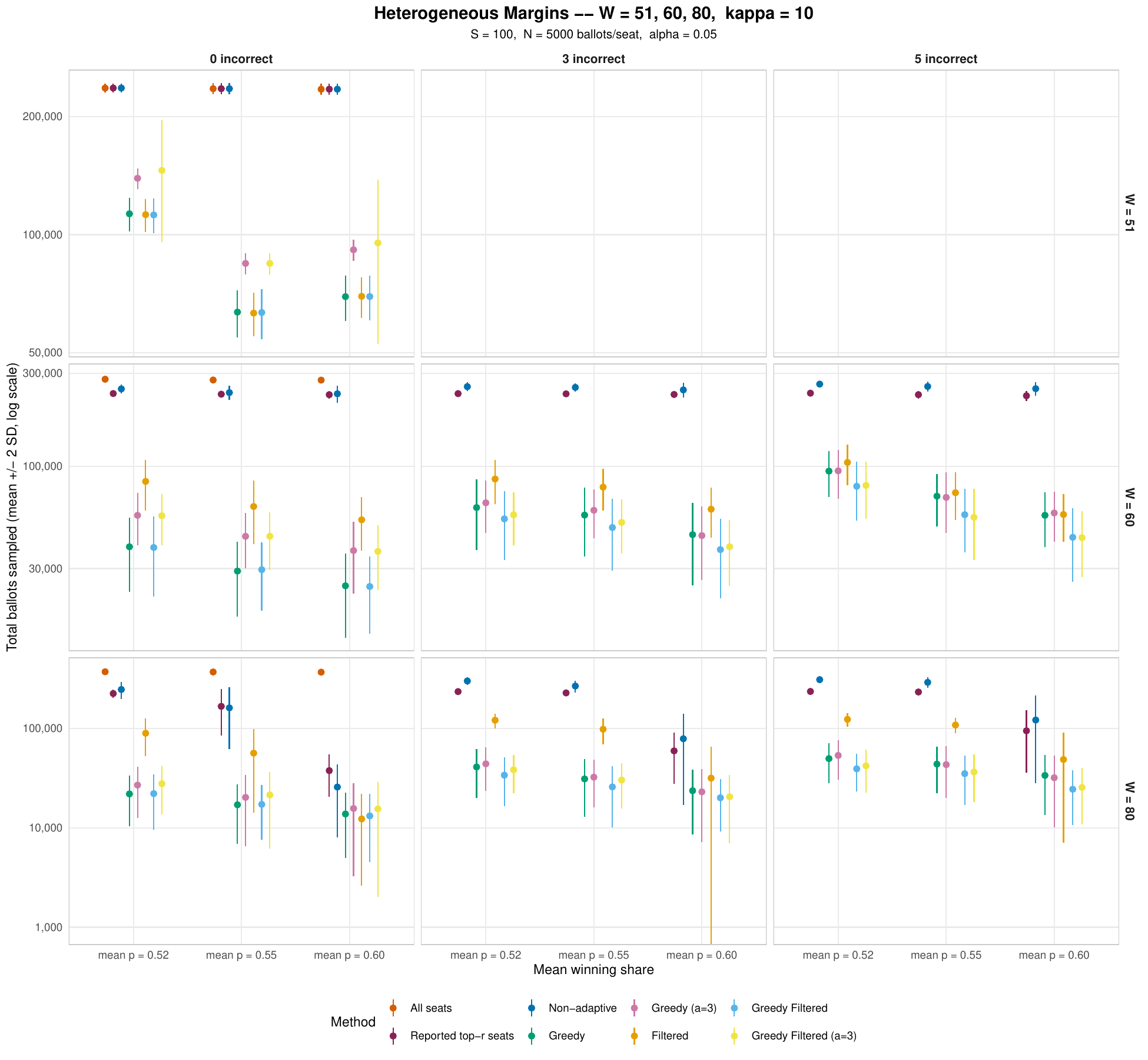}
\caption{Extended simulation results for the two-candidate plurality contests
of Scenario~2 from \autoref{sec:sim_plurality}.
The plot matches the second row of \autoref{fig:sim1_sim2}, but additionally
varies $|\cW| \in \{51, 60, 80\}$ and sets $\kappa_{\text{het}} = 10$.}
\label{fig:sim2_extended_page1}
\end{figure}

\begin{figure}[p]
\centering
\includegraphics[page=2,width=\textwidth]{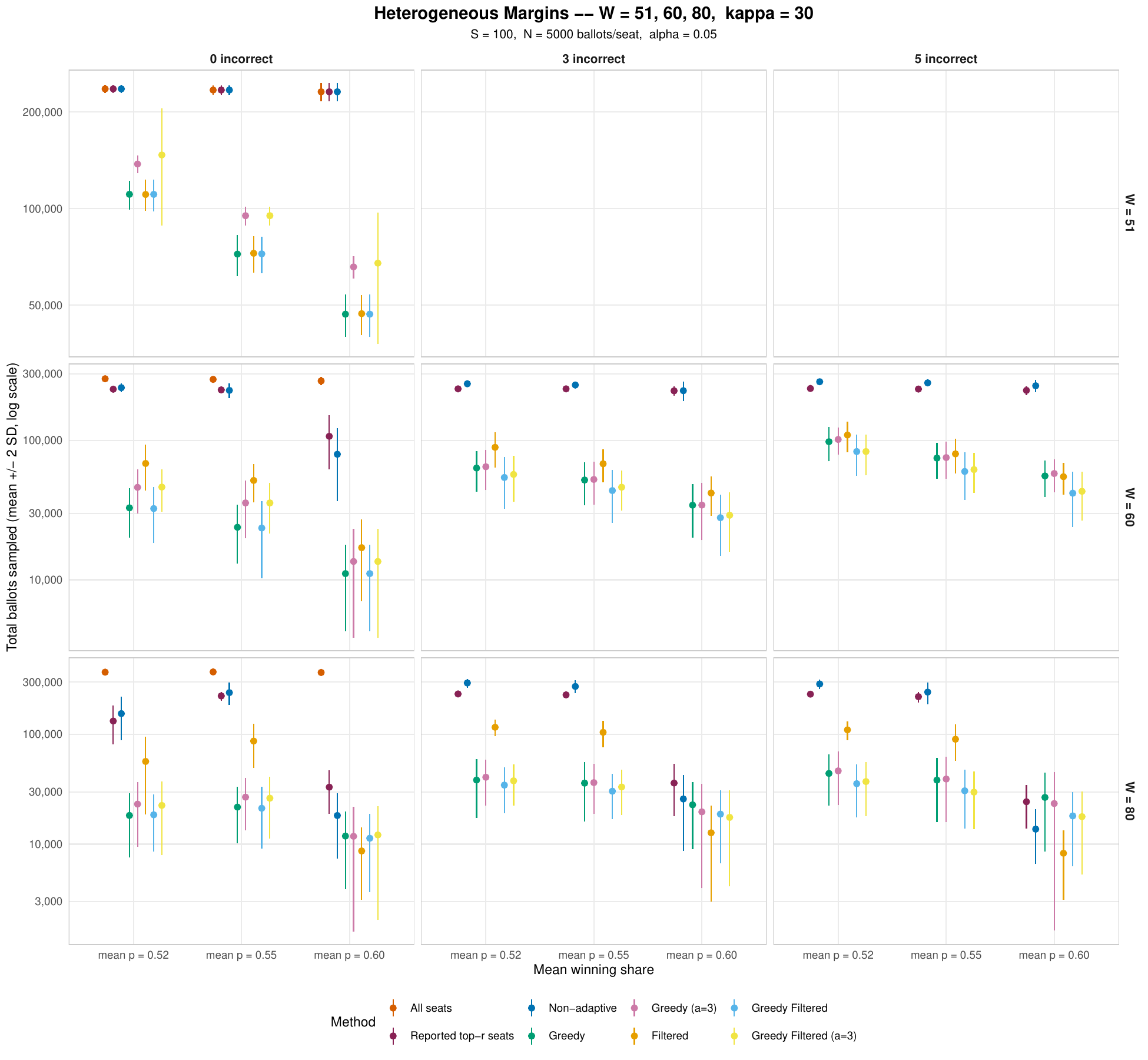}
\caption{Extended simulation results for the two-candidate plurality contests
of Scenario~2 from \autoref{sec:sim_plurality}.
The plot matches the second row of \autoref{fig:sim1_sim2}, but additionally
varies $|\cW| \in \{51, 60, 80\}$ and sets $\kappa_{\text{het}} = 30$.}
\label{fig:sim2_extended_page2}
\end{figure}

\begin{figure}[p]
\centering
\includegraphics[page=3,width=\textwidth]{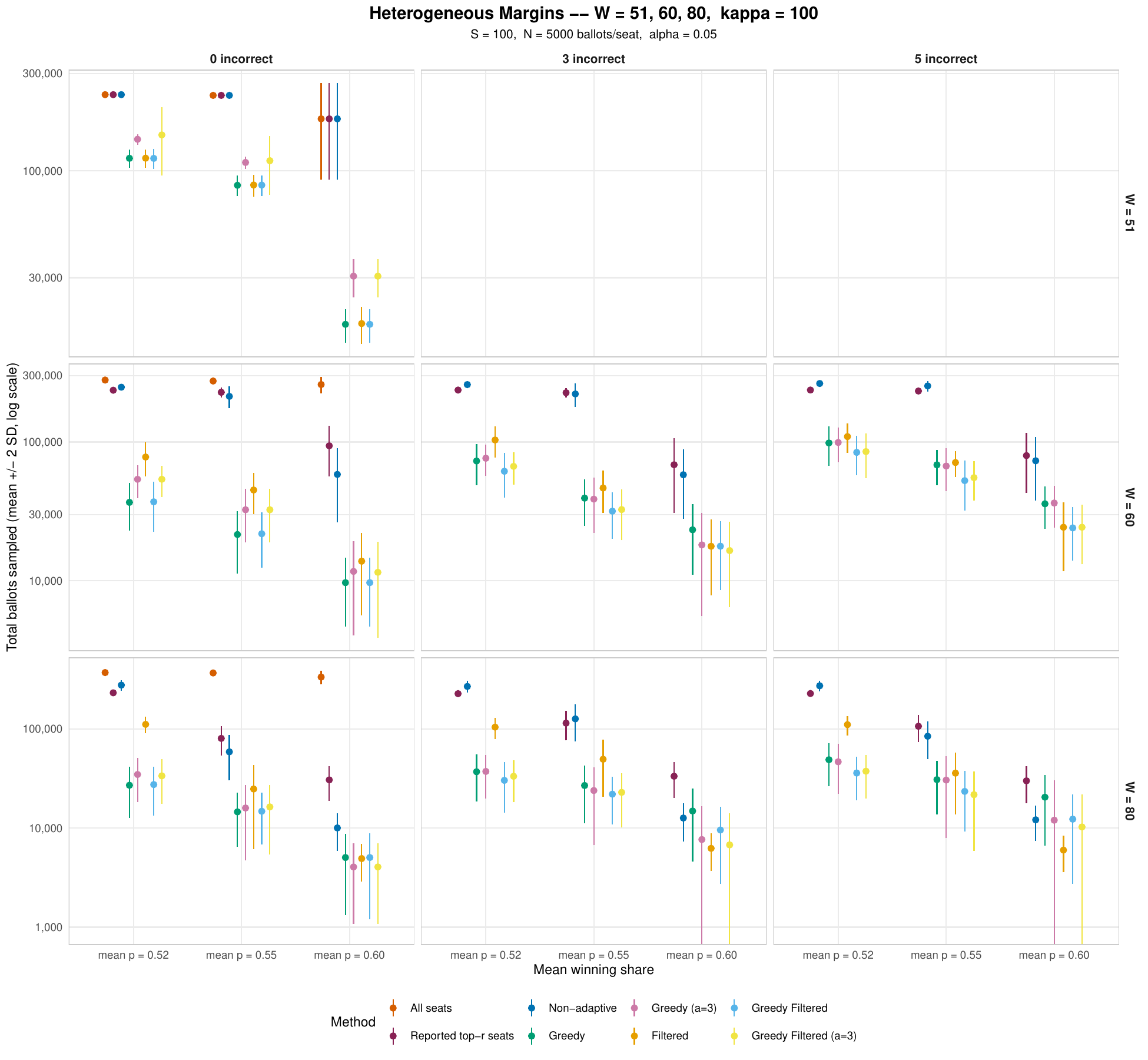}
\caption{Extended simulation results for the two-candidate plurality contests
of Scenario~2 from \autoref{sec:sim_plurality}.
The plot matches the second row of \autoref{fig:sim1_sim2}, but additionally
varies $|\cW| \in \{51, 60, 80\}$ and sets $\kappa_{\text{het}} = 100$.}
\label{fig:sim2_extended_page3}
\end{figure}

Figures \ref{fig:sim1_extended}--\ref{fig:sim2_extended_page3} show extra
simulations results beyond those presented in \autoref{sec:sim_plurality}.

\end{document}